\documentclass[manuscript]{aastex}

\usepackage{graphics,graphicx}
\usepackage{epsfig}
\usepackage{amssymb}
\usepackage{amsmath} 
\usepackage{latexsym} 

\shorttitle{Weak maser emission of methyl formate toward
  Sagittarius B2(N)} \shortauthors{Faure et al.}

\begin{document}


\title{Weak maser emission of methyl formate toward Sagittarius
  B2(N) \\ in the Green Bank Telescope PRIMOS Survey}


\author{A. Faure}
\affil{UJF-Grenoble 1 / CNRS-INSU, Institut de Plan\'etologie et
  d'Astrophysique de Grenoble (IPAG) UMR 5274, Grenoble, F-38041,
  France}

\and

\author{A.~J. Remijan} \affil{National Radio Astronomy Observatory, 520
  Edgemont Rd., Charlottesville, VA 22903, USA}

\and

\author{K. Szalewicz} \affil{Department of Physics and Astronomy,
  University of Delaware, Newark, Delaware 19716, USA}

\and

\author{L. Wiesenfeld} \affil{UJF-Grenoble 1 / CNRS-INSU, Institut de
  Plan\'etologie et d'Astrophysique de Grenoble (IPAG) UMR 5274,
  Grenoble, F-38041, France}




\begin{abstract}

A non-LTE radiative transfer treatment of {\it cis}--methyl formate
(HCOOCH$_3$) rotational lines is presented for the first time using a
set of theoretical collisional rate coefficients. These coefficients
have been computed in the temperature range 5--30~K by combining
coupled-channel scattering calculations with a high accuracy potential
energy surface for HCOOCH$_3$--He. The results are compared to
observations toward the Sagittarius~B2(N) molecular cloud using the
publicly available PRIMOS survey from the Green Bank Telescope. A
total of 49 low-lying transitions of methyl formate, with upper levels
below 25~K, are identified. These lines are found to probe a
presumably cold ($\sim 30$~K), moderately dense ($\sim
10^4$~cm$^{-3}$) and extended region surrounding Sgr~B2(N). The
derived column density of $\sim 4\times 10^{14}$~cm$^{-2}$ is only a
factor of $\sim$10 larger than the column density of the {\it trans}
conformer in the same source. Provided that the two conformers have
the same spatial distribution, this result suggests that strongly
non-equilibrium processes must be involved in their
synthesis. Finally, our calculations show that all detected emission
lines with a frequency below 30~GHz are (collisionally pumped) weak
masers amplifying the continuum of Sgr~B2(N). This result demonstrates
the importance and generality of non-LTE effects in the rotational
spectra of complex organic molecules at centimetre wavelengths.


\end{abstract}


\section{Introduction}

Interstellar methyl formate (HCOOCH$_3$) was discovered almost 40
years ago, in its most stable {\it cis}--isomeric form, in the
spectrum of the giant molecular cloud complex Sagittarius~B2
\citep{brown75,churchwell75}. Since its discovery, it has been
detected in a great variety of galactic sources including hot
molecular cores, comets and cold prestellar cores
\citep{bottinelli07,favre11,bockelee00,bacmann12,cernicharo12}. Methyl
formate is a prolate asymmetric top molecule isomeric with
acetic acid and glycolaldehyde. The two latter have also been
identified in space \citep{mehringer97,hollis00} but they are much
less ubiquitous and abundant. This result strongly suggests that the
synthesis of methyl formate is controlled by kinetic factors, since
acetic acid is thermodynamically the most stable of the three isomers
\citep{dickens01}. The chemical origin of methyl formate is however
matter of debate.

Laboratory experiments have shown that methyl formate can be formed in
the solid state from photolysis or radiolysis of methanol-bearing ices
\citep[see][and references therein]{modica12}. Gas-grain chemical
models have shown that the surface reaction between the (mobile) HCO
and CH$_3$O radicals can provide an association pathway in warm
($T\gtrsim$ 30~K) environments \citep{garrod06}. Gas-phase routes
involving methanol have also been suggested but none of these were
found to produce enough methyl formate to explain its high abundance
in hot cores ($\sim$ 10$^{-9}$-10$^{-8}$ relative to total hydrogen)
\citep{horn04}. Very recently, a new scenario of formation of complex
organic molecules was proposed by \cite{vasyunin13} to explain the
abundances observed in cold prestellar cores
(10$^{-11}$-10$^{-10}$). In this scenario, gas-phase reactions are
driven by reactive desorption of precursor species such as methanol
and formaldehyde. The predicted abundance for methyl formate is
however lower than observations by two orders of magnitude. Finally,
the higher energy {\it trans} conformational isomer of methyl formate
was recently detected toward the Sgr~B2(N) molecular cloud, suggesting
again a complex and non-equilibrium chemistry \citep{neill12}.

In all previous observational studies of methyl formate at millimetre
wavelengths, the determination of the column density was obtained by
assuming that the relative distribution of the population over all
energy levels can be described by a single Boltzmann temperature,
referred as the ``excitation'' or ``rotational'' temperature. It is
well known, however, that the low-frequency line $1_{10}-1_{11}$ at
1.61~GHz, first detected by \cite{brown75}, is weakly inverted toward
Sgr~B2. The populations of the $1_{10}$ and $1_{11}$ states are
therefore out of any thermal equilibrium in this region. This was
postulated by \cite{brown75} and evidenced by \cite{churchwell80}. In
fact, soon after the discovery of complex molecules at radio
wavelengths in the 1970's, it was accepted that the many low frequency
transitions must be inverted \citep[see][and references
  therein]{menten04}. In their pioneering work, \cite{brown75} assumed
a 1 percent inversion of the $1_{10}$ and $1_{11}$ levels of methyl
formate to derive the column density. Such a weak inversion, by
amplifying the strong background continuum radiation, is sufficient to
produce the emission of the otherwise undetectable 1.61~GHz line. The
determination of the actual inversion percentage however requires to
solve the equations of statistical equilibrium, which in turn
necessitates a good knowledge of collisional cross sections. As a
result, the accurate determination of column densities from low
frequency molecular transitions cannot rely on thermal equilibrium
analysis but requires a non-local thermodynamical equilibrium
(non-LTE) treatment. This was recognized in the early years of
radio-astronomy \citep[e.g.][]{gottlieb73}. For all complex molecules
except methanol \citep[see][and references therein]{rabli11}, however,
such treatments have been hampered so far by the lack of accurate
collisional cross sections.

Recently, rotational cross sections for methyl formate colliding with
He atoms were computed \citep{faure11}. To date, these are the first
available cross sections for an asymmetric top with eight atoms. In
the present work, the corresponding rate coefficients are deduced in
the temperature range 5--30~K. Encouraged by the availability of these
rate coefficients, we have {\it i)} searched for all centimetre wave
transitions of methyl formate toward the Sgr~B2(N) region and {\it
  ii)} performed non-LTE radiative transfer calculations. All
observational data are taken as part of the PRebiotic Interstellar
MOlecular Survey (PRIMOS). The rate coefficient calculations are
presented in Section~2. In Section~3, the PRIMOS observations are
detailed. We compare non-LTE calculations with PRIMOS data in
Section~4. Conclusions are drawn in Section~5.

\section{Rate coefficient calculations}

Quantum scattering calculations for helium interacting with methyl
formate were performed recently on the high-accuracy potential energy
surface (PES) of \cite{faure11}, where full details can be
found. Helium was employed as a substitute for H$_2$, which is five
times more abundant than He. Calculations with He are, however, much
less expensive and, in first approximation, cross sections with H$_2$
can be assumed to be equal to those for collisions with helium, as
discussed below. The interaction energies were computed at the coupled
cluster method with single, double, and noniterative triple
excitations, CCSD(T). Symmetry-adapted perturbation theory based on
the density-functional description of monomers, SAPT(DFT), was
employed to compute the asymptotic, long-range part of the PES. Methyl
formate was assumed to be rigid in its experimentally determined
geometry. The interaction energies were thus obtained on a
three-dimensional grid and they were fitted by an analytic function of
interatomic distances with correct asymptotic behavior for large
intermonomer separations. This 3D PES was then refitted using a
partial wave expansion adapted to close-coupling scattering
calculations\footnote{The fit of the PES is available upon request
  from alexandre.faure@obs.ujf-grenoble.fr.}.

\begin{figure}
\begin{center}
\includegraphics*[scale=1.0,angle=0.]{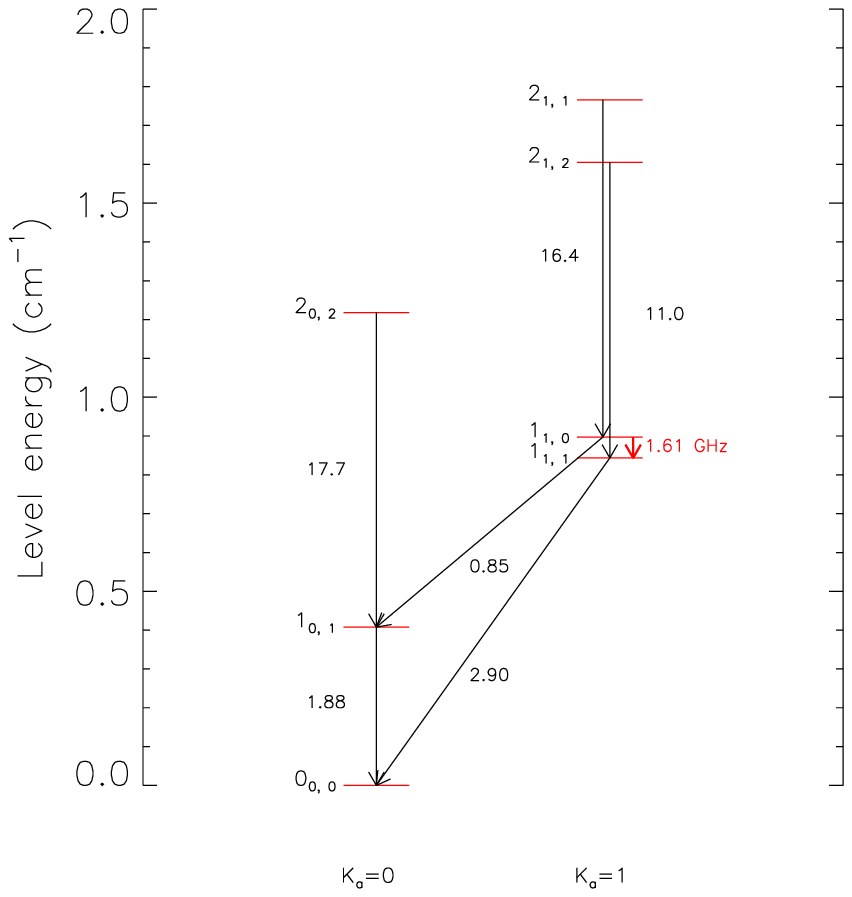}
\caption{The lowest 7 rotational levels of $A$-type HCOOCH$_3$, taken
  from the JPL catalog \citep{pickett98} available at
  www.splatalogue.net \citep{remijan07}. For each level, the strongest
  radiative transitions are indicated by arrows and the numbers give
  the corresponding Einstein coefficients (in units of
  $10^{-8}$s$^{-1}$). The 1.61~GHz line is denoted by a thick red
  arrow.}
\label{ladder}
\end{center}
\end{figure}

In the ground torsional level, each rotational level of methyl formate
is split by tunneling into a doublet, the nondegenerate $A$ level, and
the doubly degenerate $E$ level. Radiative and collisional transitions
from an $A$ sublevel to an $E$ sublevel are however not allowed. The
coupling between rotation and torsion is weak and a standard
rigid asymmetric top Hamiltonian is adequate to reproduce all levels
below the first excited torsional level (i.e. 130~cm$^{-1}$) with an
accuracy better than 0.1~cm$^{-1}$. The three rotational constants
were taken from \cite{curl59} and are listed in Table~III of
\cite{faure11}. The lowest 7 levels of $A$-type HCOOCH$_3$ are shown
in Fig.~\ref{ladder}. Scattering calculations, which are extremely
demanding owing to the dense spectrum of methyl formate, were
performed using the \texttt{OpenMP} version of the \texttt{MOLSCAT}
code\footnote{Repository at
  http://ipag.osug.fr/\~{}faurea/molscat/index.html} at the full
close-coupling level. The reduced mass of the system is
3.752371~amu. As the $A$-type and $E$-type transition energies differ
by less than 0.1~cm$^{-1}$, the cross sections for rotational
(de)excitation are very similar for the two symmetries, with
differences below 5\% \citep{faure11}. Only cross sections for the
$A$-type HCOOCH$_3$ were therefore computed and they can be employed
both for $A$-type and $E$-type HCOOCH$_3$. The coupled-channel
equations were integrated using the hybrid modified log-derivative
Airy propagator of \cite{alexander87} with a careful check of all
propagation parameters. Calculations were performed for total energies
between 0.25 and 130~cm$^{-1}$, i.e. up to the first torsional
threshold. The energy grid was adjusted to take into account the
presence of low-energy resonances. All calculations also included
several energetically closed channels: the basis set incorporated all
target states with $J\leq 14$, i.e., 225 rotational states with the
highest level, $J_{K_a, K_c}=14_{14, 0}$, at 133.5~cm$^{-1}$ above the
ground state. The cross sections were found to be converged to within
20\% (and generally within 5$-$10\%) for all transitions involving the
lowest 65 states, i.e., up to the level $9_{2, 8}$ at 19.98~cm$^{-1}$.

Rate coefficients were obtained up to $T=$30~K by integrating the
cross sections $\sigma$ over Maxwell-Boltzmann distributions of
relative velocities:
\begin{equation}
k(T)=\left(\frac{8k_BT}{\pi \mu}\right)^{1/2}\int \sigma(E_{\rm
  coll})(\frac{E_{\rm coll}}{k_BT})e^{-E_{\rm coll}/k_BT}\mathrm{d}(\frac{E_{\rm
  coll}}{k_BT}),
\end{equation}
where $\mu$ is the reduced mass of the HCOOCH$_3$-H$_2$ system,
$\mu=1.950159$, and $E_{\rm coll}=\mu v^2/2$ is the collision
energy. We thereby assume that the cross sections for He and H$_2$ are
identical and that the rate coefficients differ only for the (square
root of the) reduced mass ratio. This approximation is most valid for
the ground state of para-H$_2$($J=0$) and for heavy molecules, as
discussed in \cite{wernli07}. Larger differences are expected between
He and ortho-H$_2$($J=1$), as observed e.g. in H$_2$CO
\citep{troscompt09} and CH$_3$OH \citep{rabli11}. However, since the
ortho state ($J=1$) lies 170~K above the para state ($J=0$), it should
not be significantly populated at the temperatures investigated here
\citep[see e.g.][]{faure13}. The cross sections were extrapolated at
total energies higher than 130~cm$^{-1}$ in order to guarantee
convergence of Eq.~(1). A total of 2080 transitions were considered
for $A$-type HCOOCH$_3$, corresponding to the lowest 65 levels with
upper energies below 20~cm$^{-1}$. This complete set of
(de-)excitation rate coefficients is available online from the LAMDA
database \citep{schoier05} at
\texttt{home.strw.leidenuniv.nl/∼moldata/} and BASECOL database
\citep{dubernet13} at \texttt{www.basecol.obspm.fr}. Scattering
calculations for higher levels (and therefore temperatures) are in
progress in our group.

\begin{figure}
\begin{center}
\includegraphics*[scale=0.8,angle=-90.]{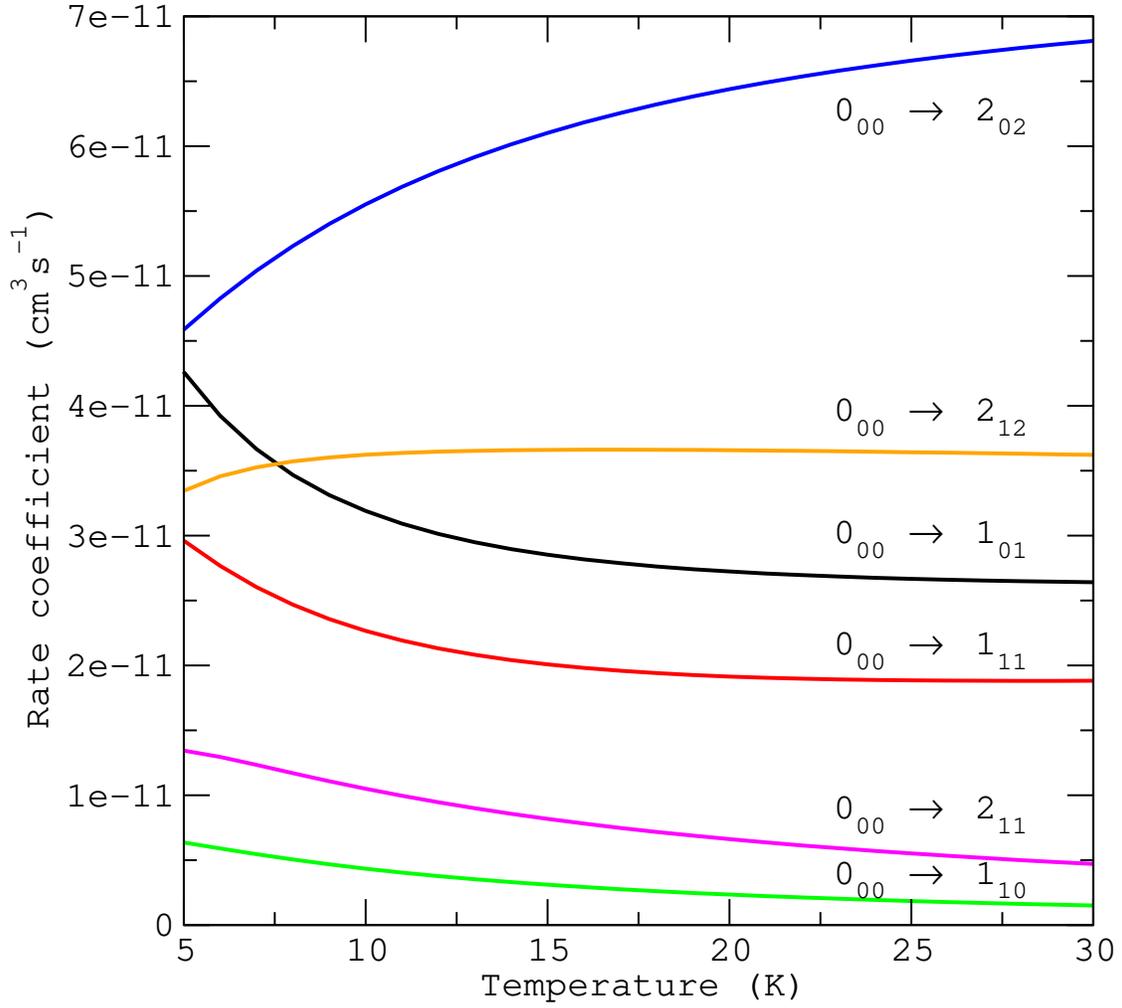}
\caption{Rate coefficients for rotational excitation out of the ground
  rotational state ($0_{00}$) of $A$-HCOOCH$_3$ as a function of
  temperature.}
\label{rate1}
\end{center}
\end{figure}

Rate coefficients for the lowest 6 transitions out of the ground state
$0_{00}$ are presented in Fig.~\ref{rate1}. We first observe that the
favored transition is $0_{00}\to 2_{02}$, corresponding to the
propensity rules $\Delta J=2$ and $\Delta K_a=0$. These propensities
reflect, respectively, the ``even'' symmetry of the PES and the
conservation of $K_a$ (the component of $J$ along the smallest moment
of inertia), as discussed in \cite{faure11}. In addition, for each
$K_a$ doublet in the $K_a=1$ ladder, the lower level (which
corresponds to $J$ preferentially oriented along the direction of the
greatest moment of inertia) is found to be favored. It should be noted
that this result was shown to explain the anti-inversion of the
doublet $1_{10}-1_{11}$ in interstellar formaldehyde
\citep{townes69,troscompt09}. In the case of methyl formate, however,
this doublet is inverted in Sgr~B2 \citep{brown75,churchwell80}. This
disparity is not surprising as the nuclear symmetries and selection
rules are different. Thus, in contrast to formaldehyde, both
collisional and radiative transitions between the $K_a=1$ and $K_a=0$
ladders are allowed in methyl formate, as detailed below.

\begin{figure}
\begin{center}
\includegraphics*[scale=0.8,angle=-90.]{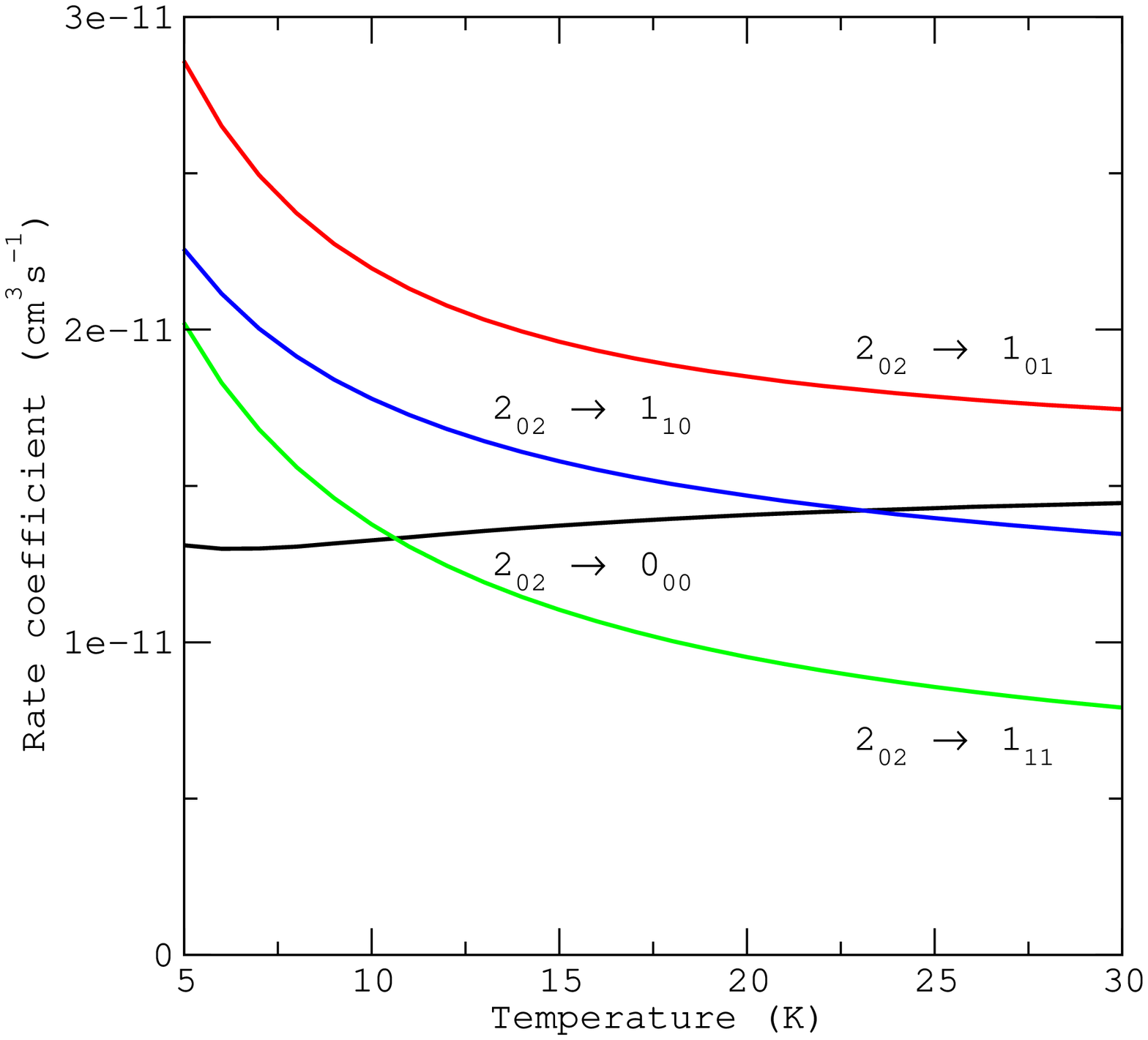}
\caption{Rate coefficients for rotational relaxation out of the
  rotational state $2_{02}$ of $A$-HCOOCH$_3$ as a function of
  temperature.}
\label{rate2}
\end{center}
\end{figure}

In Fig.\ref{rate2}, rate coefficients for collisional deexcitation
transitions from the $2_{02}$ level are presented. We note that the
largest rate is for the excitation transition $2_{02}\to 4_{04}$ (not
plotted), as expected from the above propensity rules. It is observed
in Fig.~\ref{rate2} that the level $2_{02}$ preferentially deexcites to
the level $1_{01}$. In the case of the $1_{10}-1_{11}$ doublet, the
upper level $1_{10}$ is favored with a rate which is a factor of 1.7
(at 30~K) larger than the deexcitation rate to the lower level
$1_{11}$. This contrasts with collisions from the ground state
$0_{00}$. Thus, following the initial excitation $0_{00}\to 2_{02}$,
the level $2_{02}$ relaxes preferentially to the upper level of the
doublet.

Radiatively, for an asymmetric-top with dipole components along the
$a$ and $b$ inertia axes\footnote{The $a$ and $b$ inertia axis
  correspond respectively to the $x$ and $z$ axis in the plane of the
  molecule, as defined in \cite{faure11}.}, the standard (rigorous)
selection rules are:
\begin{equation}
a{\rm -type}: \Delta J=0, \pm 1; \Delta K_a=0; \Delta K_c=\pm 1
\end{equation}
\begin{equation}
b{\rm -type}: \Delta J=0; \pm 1, \Delta K_a=\pm 1; \Delta K_c=\pm 1,
\pm 3.
\end{equation}
The strongest radiative transitions are indicated by arrows in
Fig.\ref{ladder}. It is shown in particular that the spontaneous decay
rate from the $1_{1, 1}$ to $0_{0, 0}$ level is about 3.4 times that
from the $1_{1, 0}$ to the $1_{0, 1}$ level. This argument was
employed by \cite{brown75} to explain that the doublet $1_{1, 0}-1_{1,
  1}$ is likely inverted and amplifies the continuum emission of
Sgr~B2. In fact, this radiative effect adds to the preferential
collisional deexcitation $2_{02}\to 1_{10}$. As a result, both
collisional and radiative transitions are expected to favor the
inversion of the doublet $1_{10}-1_{11}$. As the critical
densities\footnote{The critical density for a particular level is
  given by the ratio between the sum of all radiative rates and the
  sum of all deexcitation collisional rates from this level.} for the
$1_{10}$ and $1_{11}$ levels are about 300~cm$^{-3}$ and
1500~cm$^{-3}$, respectively, inversion is expected to occur at low
density ($n({\rm H_2})< 10^5$~cm$^{-3}$). Note that the above
arguments do not necessarily hold for other lines and only non-LTE
radiative transfer calculations can predict which transitions are
inverted for a given set of physical conditions.

\section{Observations}

The PRIMOS survey is a National Radio Astronomy Observatory (NRAO)
Robert C. Byrd Green Bank Telescope (GBT) Legacy
Program\footnote{Access to the entire PRIMOS data and specifics on
  observing strategies are available at
  www.cv.nrao.edu/$\sim$aremijan/PRIMOS/.} started in 2007 to provide
a complete spectral line survey in frequency ranging from
$\sim$300~MHz to $\sim$50~GHz toward the preeminent source of
molecular emission in the Galaxy - Sgr~B2(N). Of the more than 180
interstellar molecules detected, more than half have been detected
first in the Sgr~B2 star-forming region. This region contains compact
hot molecular cores of arcsecond dimensions
\citep{belloche08,belloche09,nummelin00,liu99}, molecular maser
emitting regions \citep[see e.g.][]{mcgrath04,
  mehringer94,mehringer97,gaume90}, and ultracompact continuum sources
surrounded by larger-scale continuum features as well as molecular
material extended on the order of arcminutes \citep[see e.g.][and
  references therein]{jones11}.  In addition, small-scale and
large-scale shock phenomena characterize the complex. In particular,
the hot molecular core known as the ``Large Molecule Heimat'' (LMH)
has for the last twenty years been the first source searched to detect
and identify new large interstellar molecules since many of the large
organic species have previously been confined to its $\sim$~5''
diameter \citep{miao97}. However, the recent GBT detections of large
organic molecules \citep[][and references
  therein]{neill12,zaleski13,loomis13}, have suggested that prebiotic
molecules found toward the Sgr~B2(N) complex are extended, perhaps
even on the order of a 2'x2' field. Data were taken in the OFF-ON
position switching mode with two-minute scans toward the ``ON''
pointing position ($\alpha$J2000 = 17h47m19.8s, $\delta$J2000 =
$-28^{\rm o}22'17.0"$). Observations continue to expand the frequency
range of the PRIMOS survey. For more details on the PRIMOS survey
observations see \cite{neill12}.

\begin{figure}
\begin{center}
\includegraphics*[scale=0.8,angle=-90.]{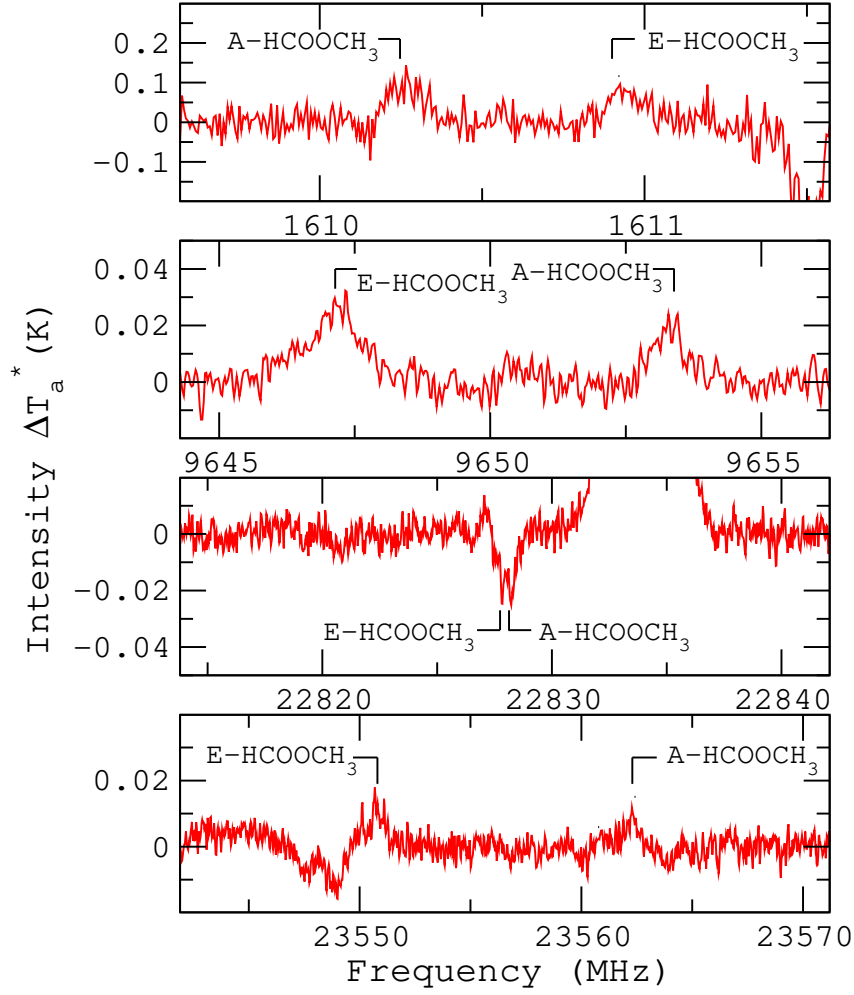}
\caption{Spectral features of $A$- and $E$-methyl formate towards
  Sgr~B2(N) from PRIMOS data. From top to bottom: the $1_{10}-1_{11}$
  emission lines at 1610.2 and 1610.9~MHz, the $3_{12}-3_{13}$
  emission lines at 9647.1 and 9653.4~MHz, the $2_{12}-1_{11}$
  absorption lines at 22827.7 and 22828.1~MHz and the $4_{13}-4_{04}$
  emission lines at 23550.8 and 23562.3~MHz.}
\label{data}
\end{center}
\end{figure}

From 300 MHz to 50~GHz, a total of 49 methyl formate transitions with
upper levels below 25~K were firmly identified (23 $A$-type and 26
$E$-type). We note that about 80\% of the detected lines have upper
levels below 10~K. Examples of spectral features are illustrated in
Fig.~\ref{data} for both emission and absorption lines. An LSR source
velocity of $+$64~km$\cdot$s$^{-1}$ was assumed. Typical line widths
are $\sim$10~km$\cdot$s$^{-1}$. The peak intensities were measured at
each observed transition and are reported below in Figs.~\ref{nonlte1}
and \ref{nonlte2}. The authors encourage the reader to obtain the full
set of spectral line data from the PRIMOS website, in order to perform
more detailed analyses.

\section{Non-LTE calculations}

Radiative transfer calculations were performed with the \texttt{RADEX}
code \citep{vandertak07}, using the Large Velocity Gradient (LVG)
approximation for a uniform expanding sphere. The \texttt{RADEX} LVG
code was employed to compute the line excitation temperatures
($T_{ex}$) and opacities ($\tau$) for a given column density ($N_{\rm
  tot}$), kinetic temperature ($T_{\rm kin}$) and density of H$_2$
($n({\rm H_2})$). The two symmetries $A$-HCOOCH$_3$ and $E$-HCOOCH$_3$
were treated separately and we included 65 rotational levels for both
species, with the highest energy level ($9_{2, 8}$) lying at 28.8~K
above the ground state. Grids of models were computed for kinetic
temperatures of 5--30~K (as constrained by the collisional data),
column densities between 10$^{12}$ and $10^{16}$cm$^{-2}$ and H$_2$
densities in the range 10--10$^7$~cm$^{-3}$. The line width (FWHM) was
fixed at 10.0~kms$^{-1}$, as deduced from the PRIMOS spectrum.

For each line in the range 1--49~GHz, the solution of the radiative
transfer equation was expressed as the antenna temperature:
\begin{equation}
\Delta T_a^*=[J_\nu(T_{ex})-J_\nu(T_{cmb})-T_{c}](1-e^{-\tau}),
\label{tmb}
\end{equation}
where $J_\nu(T)=(h\nu/k_B)/(e^{h\nu/k_BT}-1)$, $\nu$ is the frequency
of the transition, $T_{ex}$ and $\tau$ are the excitation temperature
and opacity computed by \texttt{RADEX}, $T_{cmb}=$2.725~K is the
temperature of the (local) cosmic microwave brackground and $T_{c}$ is
the main beam background continuum temperature of Sgr~B2(N), as
measured by the GBT, which varies from $\sim$95~K at 1.6~GHz down to
$\sim$1.2~K at 49~GHz. This continuum is consistent with optically
thin, nonthermal (synchrotron) emission with a flux density spectral
index of -0.7 and a gaussian source size of $\sim$143'' at 1~GHz that
decreases with increasing frequency as $\nu^{-0.52}$ \citep{hollis07}.

In Eq.~\ref{tmb}, it is assumed that the methyl formate cloud is
extended and fills the GBT half-power beamwidth ($\theta_B\approx
740''/\nu$, where $\nu$ is in units of GHz) with a unit filling
factor\footnote{It is also implicitly assumed that the continuum
  source is entirely covered by the HCOOCH$_3$ cloud.}. This
approximation is reasonable for transitions above $\sim$5~GHz since
many complex molecules like, e.g., glycolaldehyde have significant
spatial scales on the order of arcminutes \citep{hollis04}. At lower
frequencies, the hypothesis of a unit filling factor is
questionable. However, in this frequency range, both $|J_\nu(T_{ex})|$
and $J_\nu(T_{cmb})$ are much smaller than $T_{c}$ (see below) so that
beam dilution effects are small and Eq.~\ref{tmb} reduces to:
\begin{equation}
\Delta T_a^*\approx -T_{c}(1-e^{-\tau}),
\label{invert}
\end{equation}
and for an optically thin line:
\begin{equation}
\Delta T_a^*\approx -T_{c}\tau.
\label{invert2}
\end{equation}
It should be noted that according to Eq.~\ref{invert}, a low frequency
transition will be observed in emission only if it is inverted
(i.e. has a negative opacity), otherwise it will be in absorption.

\begin{figure}
\begin{center}
\includegraphics*[scale=0.8,angle=-90.]{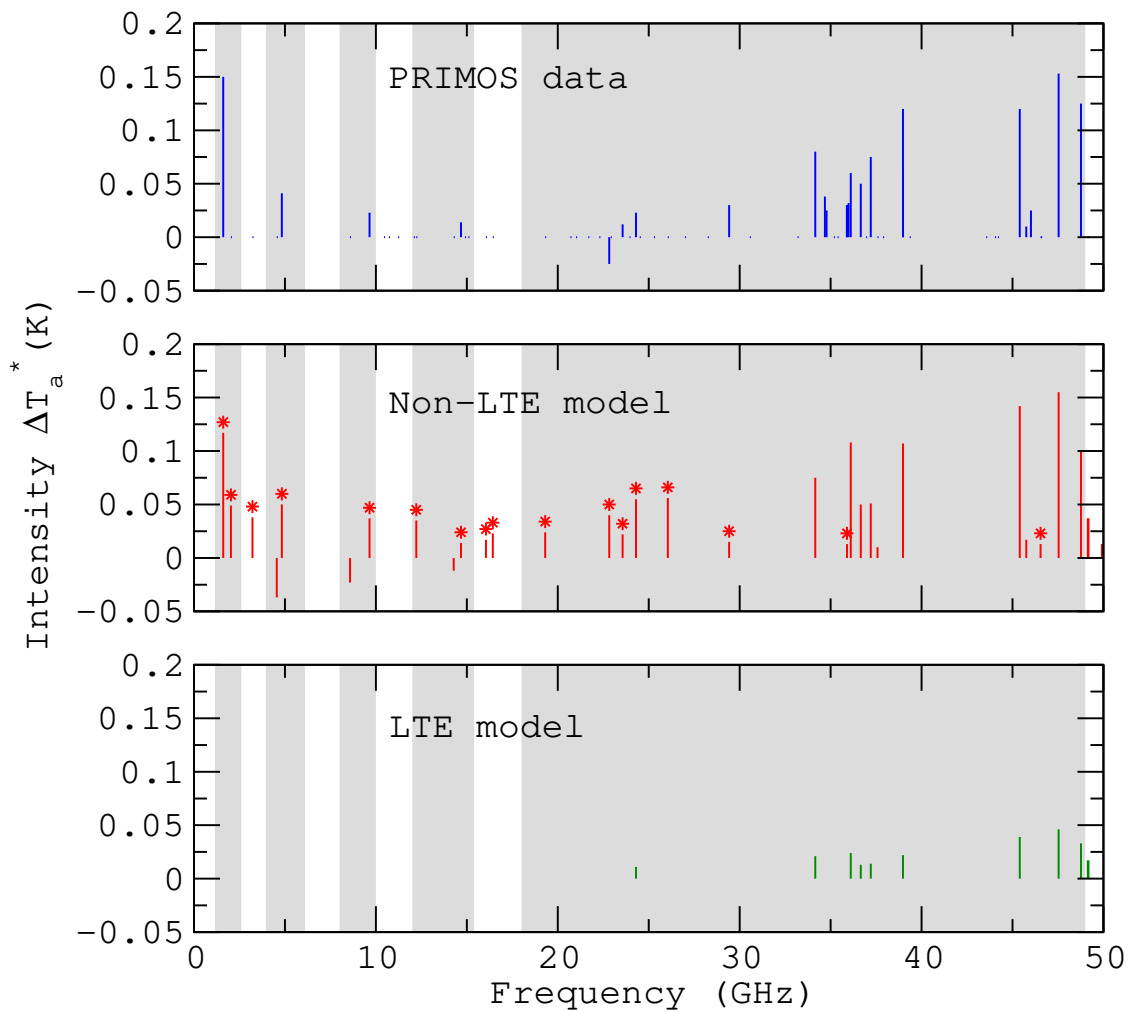}
\caption{Intensity ($\Delta T_a^*$ in K) of $A$-methyl formate
  rotational transitions in the frequency range 0-50~GHz. PRIMOS data
  are plotted in the upper panel. Our best fit non-LTE results are
  represented in the middle panel. All lines marked with an asterisk
  are weak masers amplifying the continuum of Sgr~B2(N). The bottom
  panel gives LTE results. The shaded areas show the observing
  passbands in the PRIMOS survey.}
\label{nonlte1}
\end{center}
\end{figure}
\begin{figure}
\begin{center}
\includegraphics*[scale=0.8,angle=-90.]{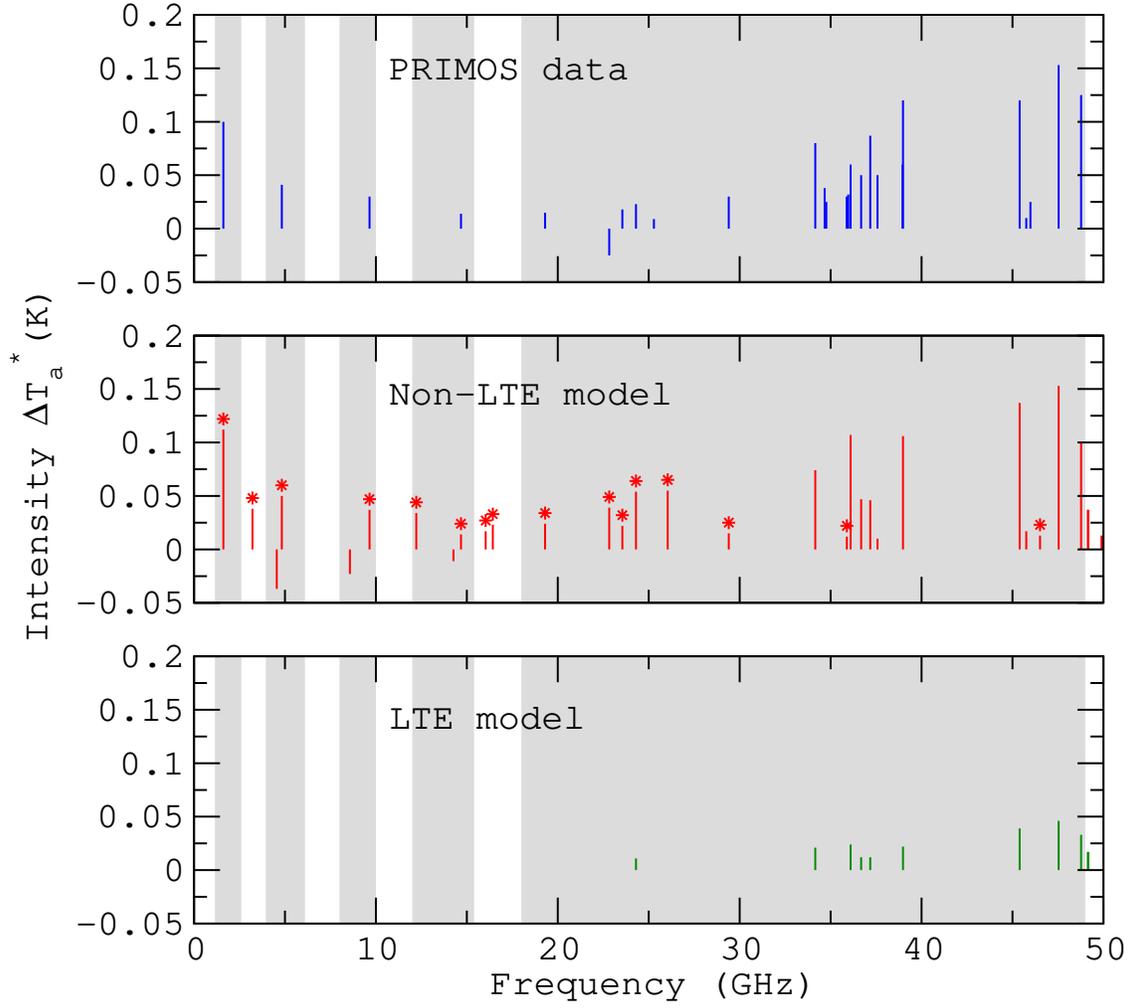}
  \caption{Same as Fig.~\ref{nonlte1} for $E$-methyl formate.}
\label{nonlte2}
\end{center}
\end{figure}

An unweighted least-squares fit was performed to determine the
physical parameters that best reproduce the emission intensities in
the PRIMOS spectrum. All 49 detected transitions of methyl formate
were employed in the minimization procedure. A good agreement was
found between predicted and observed values, as shown in
Figs.~\ref{nonlte1} and \ref{nonlte2} where line intensities from our
best non-LTE model are compared to the PRIMOS data in the frequency
range 1--49~GHz. Lines with computed intensities below 10~mK are not
displayed. It is observed that the intensities of the 10 strongest
lines (those above 100~mK) are reproduced to within 25\% by our
non-LTE model. The average error between the data and the model is
20~mK. We also notice that the 2 PRIMOS absorption lines at 22.83~GHz
(displayed in Fig.~\ref{data}) are predicted in emission by our
model. Inversely, a few lines are predicted in absorption in our model
but they are not detected. As discussed below, these differences
likely reflect density and temperature inhomogeneities. We also note
that some predicted lines are not observed in the PRIMOS spectrum
owing to contamination by other species, e.g. the $A$-type transition
at 19.3~GHz which coincides with a recombination line.

The physical parameters from the best fit model are listed in
Table~\ref{best}. We notice that they are identical for $A$-HCOOCH$_3$
and $E$-HCOOCH$_3$, as expected. As a result, the $A$/$E$ ratio is
found to be equal to unity and the total column density of HCOOCH$_3$
is $4\times 10^{14}$~cm$^{-2}$. The goodness of the fit can be judged
by the reduced chi-squared values of 1.8--1.9 and the corresponding
root mean square deviations of 27~mK. The fit is thus surprisingly
good given the simplicity of the physical model employed (uniform
sphere) compared to the complexity of the Sgr~B2 source. Error limits
on the physical parameters are difficult to estimate, in particular
the actual kinetic temperature might exceed 30~K which is the highest
available value from our collisional data. This important point is
further discussed in Section~5. Uncertainties in the parameters
$n({\rm H_2})$ and $N_{\rm tot}$ were estimated from the analysis of
$\chi^2_{\rm red}$ confidence contour levels and are typically a
factor of 2. As shown below, however, the lowest-frequency transitions
are much more sensitive to the density of H$_2$ than the
highest-frequency ones.

\begin{table}
\caption{Physical parameters of Sgr~B2(N) as determined from the best
  fit \texttt{RADEX} model of the HCOOCH$_3$ PRIMOS data. The reduced
  chi-squared ($\chi^2_{\rm red}$) and the root mean square deviation
  (rms) of the best fit are also provided.}
\label{best}
\begin{center}
\begin{tabular}{lcc}
\hline\hline
                           & $A$-HCOOCH$_2$   & $E$-HCOOCH$_3$\\ \hline
$T_{\rm kin}$ (K)           & 30               & 30 \\ 
$n({\rm H_2})$ (cm$^{-3}$) & $1.3\times 10^4$ & $1.3\times 10^4$ \\ 
$N_{\rm tot}$ (cm$^{-2}$)   & $2.0\times 10^{14}$ & $2.0\times 10^{14}$ \\ 
$\chi^2_{\rm red}$         & 1.84             & 1.88 \\ 
rms (mK)                  & 27.1              & 27.4  \\                   
\hline\hline
\end{tabular}
\end{center}
\end{table}

We have also reported in Figs.~\ref{nonlte1} and \ref{nonlte2} (bottom
panels) the predicted spectrum for LTE conditions where all HCOOCH$_3$
lines are thermalized at the kinetic temperature, i.e. $T_{\rm
  kin}=30$~K, keeping the column density fixed at the best fit
value. We observed that all line intensities are significantly
underestimated, especially those at the lowest frequencies. This plot
thus clearly illustrates the non-LTE nature of the GBT methyl formate
spectrum.

\begin{figure}
\begin{center}
\includegraphics*[scale=0.8,angle=-90.]{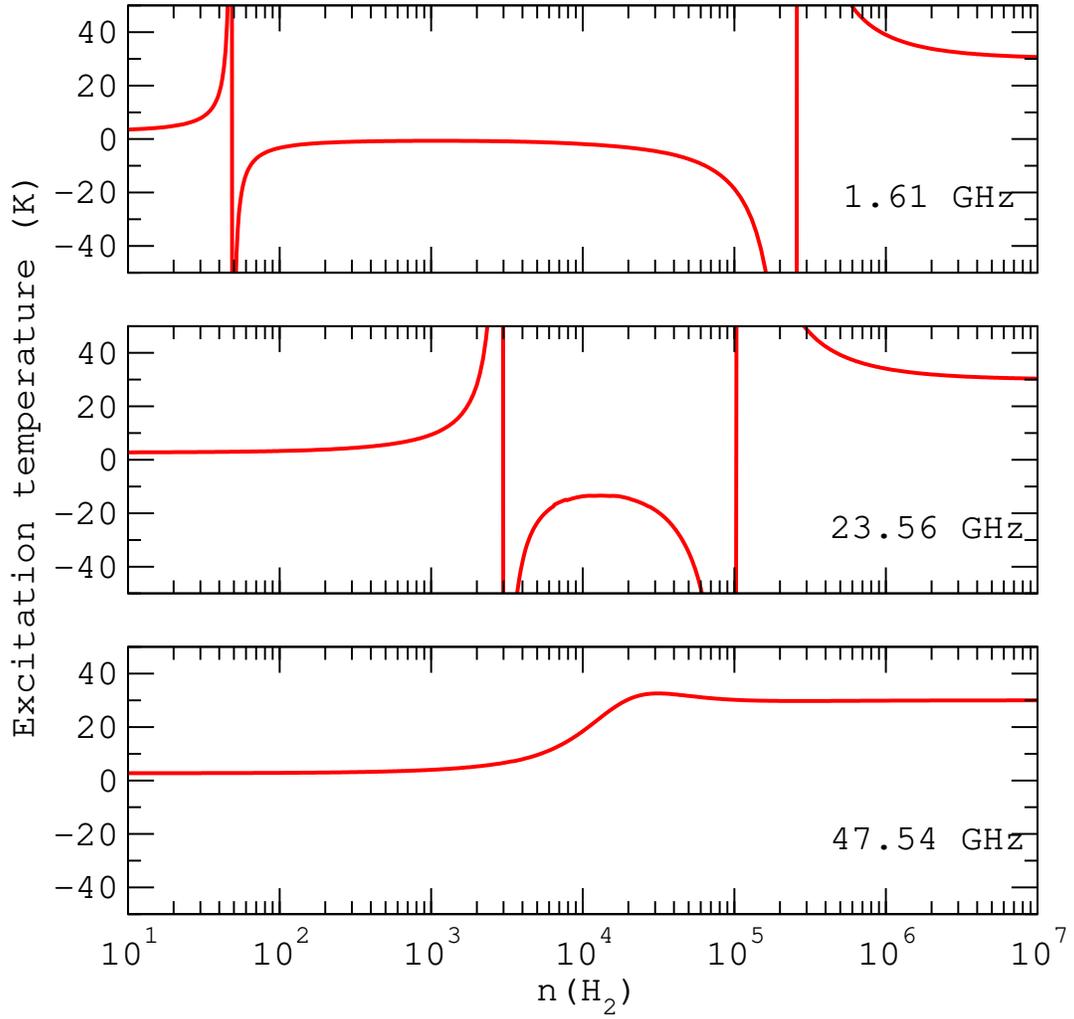}
\caption{Excitation temperature of $A$-HCOOCH$_3$ rotational
  transitions at 1.61~GHz ($1_{10}-1_{11}$), 23.56~GHz
  ($4_{13}-4_{04}$) and 47.54~GHz ($4_{04}-3_{03}$) as a function of
  H$_2$ density.}
\label{tex}
\end{center}
\end{figure}

In Figs.~\ref{nonlte1} and \ref{nonlte2}, all transitions marked with
an asterik are (weak) masers, i.e. their excitation temperature and
opacity are negative. We thus observe that {\it all detected} emission
lines with a frequency below 30~GHz are masers. Inversion was of
course expected for the lowest 1.61~GHz transition, as anticipated by
\cite{brown75}. These authors employed Eq.~\ref{invert2} above, from
which they derived the (negative) apparent opacity $\tau=-\Delta
T_a^*/T_c$. They chose an arbitrary 1 percent inversion\footnote{The
  population inversion precentage is defined as $(n_u-n_l)/(n_u+n_l)$
  where $n_u$ and $n_l$ are the populations of the upper and lower
  level of the transition, respectively.} for the $1_{10}-1_{11}$
doublet, corresponding to an excitation temperature $T_{ex}=-3.8$~K,
and they derived a column density of $1.8\times 10^{13}$~cm$^{-2}$ for
the lower $A$-state. The excitation temperature computed with
\texttt{RADEX} for the 1.61~GHz line of $A$-HCOOCH$_3$ is plotted in
the top panel of Fig.~\ref{tex}. In this figure, the kinetic
temperature is fixed at 30~K and the column density at $2\times
10^{14}$~cm$^{-3}$. The excitation temperature of the 1.61~GHz line is
found to be negative over a wide range (4 orders of magnitude) of
densities: $n({\rm H_2})$ from 50 to $3\times 10^5$~cm$^{-3}$.  We
thus demonstrate that the inversion of the $1_{10}-1_{11}$ doublet is
a robust phenomenon. In our best model ($n({\rm H_2})=1.3\times
10^4$~cm$^{-3}$), the excitation temperature is -2.3~K, corresponding
to $\sim$1.7 percent inversion, which is close to the guessed value of
\cite{brown75}.

The density range where inversion occurs is however much shorter for
other transitions. For example, the line at 23.56~GHz (medium panel in
Fig.~\ref{tex}) is inverted over the range $n({\rm H_2})$ from
$3\times 10^3$ to $10^5$~cm$^{-3}$, with $T_{ex}=$-13.4~K in our best
model (i.e. $\sim$4 percent inversion). In fact, this density range
tends to decrease with increasing frequency and above 30~GHz, most
transitions cannot be inverted. For these lines, the excitation
temperature approaches the kinetic temperature of 30~K
(i.e. thermalization) above a few $10^4$~cm$^{-3}$. An example is
given with the line at 47.54~GHz plotted in the bottom panel of
Fig.~\ref{tex}. We note that the critical densities are typically
$\sim 10^3-10^4$~cm$^{-3}$ for all levels below 30~K. Thermalization
of the rotational lines is therefore expected at densities above
10$^5$~cm$^{-3}$, as shown in Fig.~\ref{tex}, but the occurence of
population inversion is difficult to predict as it is very sensitive
to the collisional rates. We note in particular that non-inverted
transitions should appear in absorption against the continuum, as
predicted by our model for the lines at 4.55, 8.57 and 14.27~GHz (see
Figs.~\ref{nonlte1} and \ref{nonlte2}). These lines are however not
detected. The only detected absorption lines are those at 22.83~GHz
($2_{12}-1_{11}$), which are predicted in emission by our best
model. These discrepancies might simply reflect density and
temperature inhomogeneities in front of Sgr~B2(N). Indeed, at
densities higher than $\sim 5\times 10^4$~cm$^{-3}$, the predicted
absorption lines disappear (their intensities fall below
10~mK). Inversely, the population inversion of the 22.83~GHz
transitions is quenched at densities lower than $\sim 5\times
10^3$~cm$^{-3}$ where the two lines are predicted in absorption.

\begin{figure}
\begin{center}
\includegraphics*[scale=0.8,angle=-90.]{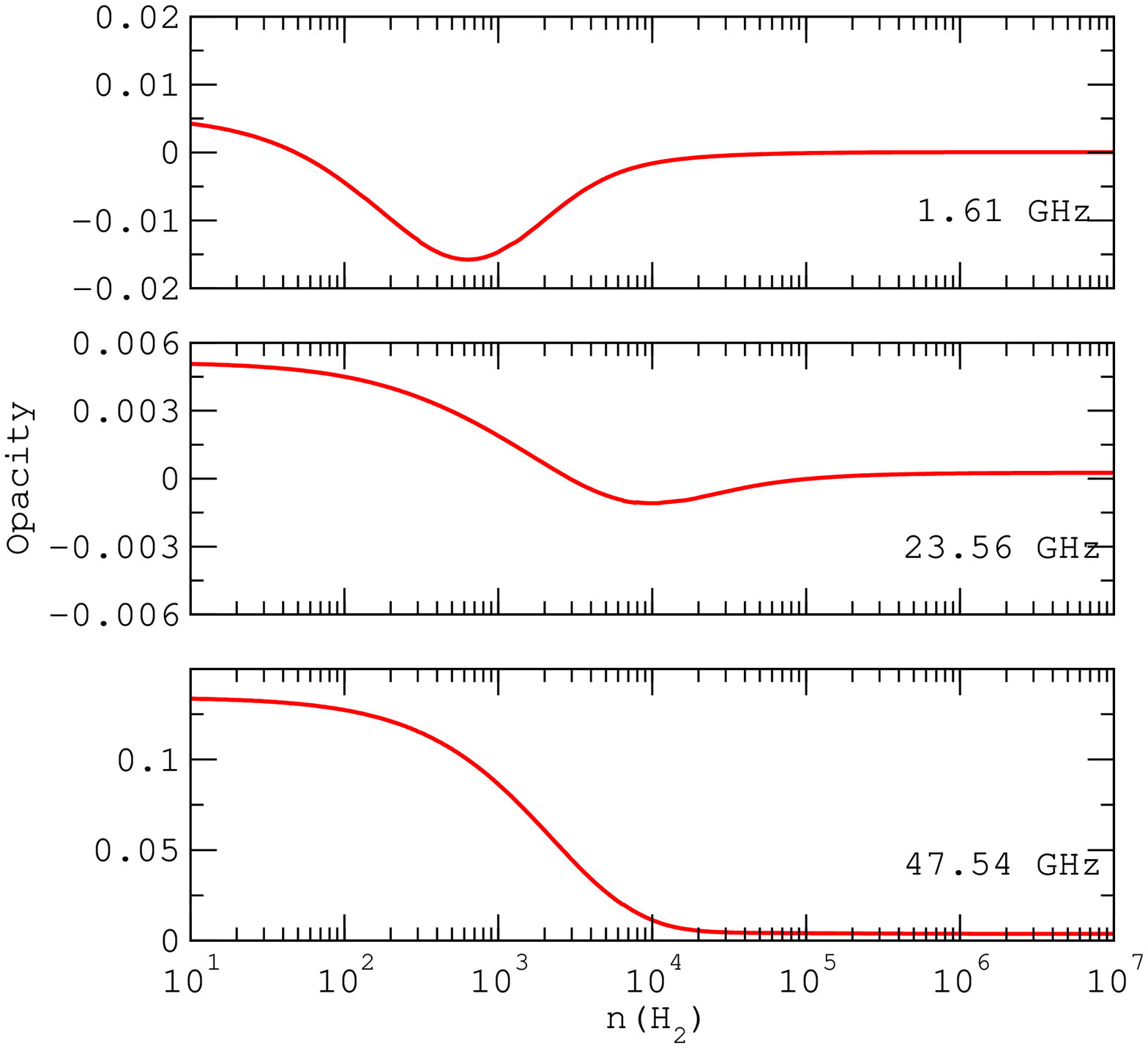}
\caption{Opacity of $A$-HCOOCH$_3$ rotational transitions at 1.61~GHz
  ($1_{10}-1_{11}$), 23.56~GHz ($4_{13}-4_{04}$) and 47.54~GHz
  ($4_{04}-3_{03}$) as a function of H$_2$ density. The kinetic
  temperature and column density of $A$-HCOOCH$_3$ are 30~K and
  $2.0\times 10^{14}$~cm$^{-2}$, respectively.}
\label{opa}
\end{center}
\end{figure}

The opacities of the three above transitions (at 1.61, 23.56 and
47.54~GHz) are plotted in Fig.~\ref{opa}. For the two maser lines, the
largest opacities (in absolute value) are found close to the critical
densities of the corresponding levels, i.e. $\sim 10^3$~cm$^{-3}$ for
the line at 1.61~GHz and $\sim 10^4$~cm$^{-3}$ for the line at
23.56~GHz. For the non-inverted transition at 47.54~GHz, the opacity
is found to increase significantly with decreasing H$_2$ density below
$\sim $10$^4$~cm$^{-3}$. In contrast to the lowest frequency
(1--5~GHz) transitions, where the line intensity depends only on $T_c$
and $\tau$ (see Eq.~\ref{invert}), the three terms $J_\nu(T_{ex})$,
$J_\nu(T_{cmb})$ and $T_{c}$ are in competition at higher
frequencies. In this regime the value of the excitation temperature is
crucial.

\begin{figure}
\begin{center}
\includegraphics*[scale=0.8,angle=-90.]{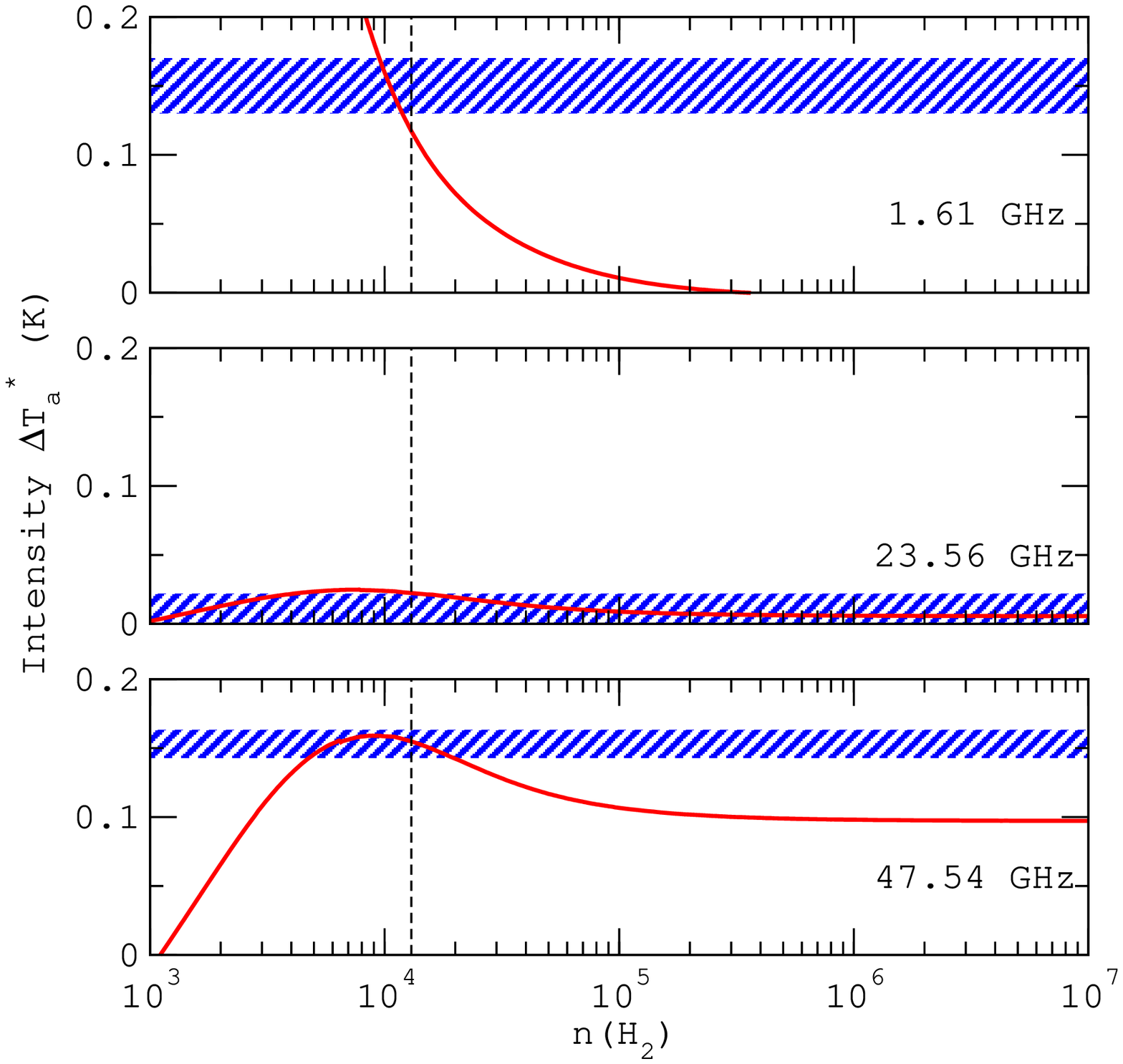}
\caption{Antenna temperatures $\Delta T_a^*$ of $A$-HCOOCH$_3$
  rotational transitions at 1.61~GHz ($1_{10}-1_{11}$), 23.56~GHz
  ($4_{13}-4_{04}$) and 47.54~GHz ($4_{04}-3_{03}$) as a function of
  H$_2$ density. The GBT intensities are represented by the blue
  hatched zone. The dashed vertical lines denote the density of the
  best model, $n({\rm H_2})=1.3\times 10^4$~cm$^{-3}$. The kinetic
  temperature and column density of $A$-HCOOCH$_3$ are 30~K and
  $2.0\times 10^{14}$~cm$^{-2}$, respectively.}
\label{ant}
\end{center}
\end{figure}

This point is clearly illustrated in Fig.~\ref{ant} where the
intensities of the three lines are found to peak at about the critical
densities (the 1.61~GHz line has a maximum of $\sim$1.5~K at $n({\rm
  H_2})\sim$10$^3$~cm$^{-3}$). Finally, Fig.~\ref{ant} also shows that
the GBT intensities of the three lines (blue hatched zones) are
reproduced for $n({\rm H_2})\sim 10^4$~cm$^{-3}$, as expected from the
best fit results. We also notice that the 1.61~GHz line gives the
strongest constraint on the H$_2$ density.

\section{Discussion}

The above non-LTE calculations have demonstrated that the GBT methyl
formate observations are sampling a presumably cold ($T_{\rm kin}\sim
30$~K) and moderately dense ($n({\rm H_2})\sim 10^4$~cm$^{-3}$)
extended region surrounding Sgr~B2(N). This result is consistent with
recent PRIMOS studies showing that many of the large molecules
identified toward Sgr~B2(N) have cold rotational temperatures
($\sim$10~K) \cite[see][and references therein]{zaleski13}. This is
also in agreement with older millimetre-wave studies of methyl formate
in Sgr~B2. Thus \cite{churchwell80} and \cite{cummins86} found
rotational temperatures of $\sim$8 and $23\pm 4$~K, respectively, from
HCOOCH$_3$ lines between 70 and 150~GHz (and energy levels below
60~K). These rotational temperatures are below our estimate of the
kinetic temperature, as expected for subthermal excitation. The methyl
formate column density was also estimated by these authors who found
$N_{\rm tot}\sim 2.8\times 10^{14}$~cm$^{-2}$ and $3.4\pm 0.3\times
10^{14}$~cm$^{-2}$, respectively. Our determination of the column
density ($\sim 4\times 10^{14}$~cm$^{-2}$) is thus in good agreement
with these pioneering studies. As noted above, however, the kinetic
temperature could exceed 30~K, which is the highest available
temperature in our collisional data. We note in this context that high
energy transitions of ammonia and other tracers observed in absorption
toward Sgr~B2 were interpreted as an evidence for a hot layer with
$T_{\rm kin} \sim 200-1000$~K, outside the cold ($20-40$~K) envelope
\citep{huttemeister95,ceccarelli02}. The present model does not firmly
exclude such high temperatures. However, with critical densities of
$\sim 10^3-10^4$~cm$^{-3}$, the rotational temperatures of methyl
formate and other heavy molecules ($\sim 10-20$~K) are not expected to
differ substantially from the kinetic temperature. It seems therefore
unlikely that the low-energy transitions of methyl formate originate
from a hot gas.

On the other hand, our results are in sharp contrast with
interferometric observations showing that HCOOCH$_3$ resides
predominantly in the compact ($\leq 5$'') LMH hot core near Sgr~B2(N)
\citep{miao95}. This source is indeed characterized by very high
densities ($>10^7$~cm$^{-3}$) and gas temperatures
($>100$~K). Assuming a source size of 4'' and a rotational temperature
of 80~K, a column density of 4.5$\times 10^{17}$~cm$^{-2}$ was derived
recently for HCOOCH$_3$ in the LMH \citep{belloche09,belloche13},
which is three orders of magnitude larger than the value inferred in
the present work. The abundance of methyl formate with respect to
H$_2$ in the LMH was estimated by \cite{belloche09} as $3.5\times
10^{-8}$, assuming an H$_2$ column density of 1.3$\times
10^{25}$cm$^{-2}$. It is difficult to have a good determination of the
H$_2$ column density in the region of cold HCOOCH$_3$
emission. However, this region was probed by several molecular tracers
in the past and rough estimates were inferred. Thus, using continuum
emission, CO isotope emission as well as several transitions of
HC$_3$N, \cite{lis91} determined a H$_2$ column density of $\sim
10^{24}$~cm$^{-2}$ for the envelope of Sgr~B2(N) characterized by a
mean density of $\sim 10^5$~cm$^{-3}$ and a kinetic temperature of
$\sim 20-40$~K. The methyl formate abundance in the cold region
surrounding Sgr~B2(N) can thus be estimated as $\sim 4\times
10^{-10}$, which is two orders of magnitude less than in the LMH hot
core and very similar to the abundances derived by \cite{bacmann12} in
the cold ($\lesssim 10$~K) protostellar core L1689B. This result also
supports low kinetic temperatures.

The detection of methyl formate and other complex molecules in the
cold interstellar gas ($T_{\rm kin}\lesssim 30$~K) challenges our
current understanding of interstellar chemistry. Indeed, reactive
heavy radicals like HCO and CH$_3$O cannot easily diffuse on cold
grain surfaces. Non-thermal processing of the grain mantles, such as
cosmic-ray bombardment or ultraviolet radiation, are therefore
required to induce diffusion, reaction and
evaporation. Post-evaporation gas-phase chemistry may also be crucial
in producing larger species from precursors formed on grains like
formaldehyde and methanol. An interesting constraint on the chemistry
of methyl formate is possibly provided by the recent (tentative)
detection of {\it trans}-methyl formate {\it in absorption} toward
Sgr~B2(N) using the PRIMOS survey \citep{neill12}. These authors have
determined a column density of $3.1\pm 1.2 \times 10^{13}$~cm$^{-2}$
and a rotational temperature of $7.6\pm 1.5$~K, suggesting that the
{\it trans} conformer also traces the cold envelope surrounding
Sgr~B2(N). Assuming the two conformers have the same spatial
distribution, the {\it cis/trans} ratio is thus found to be $\sim$10
in the cold gas, indicating strongly non-equilibrium
processes. Indeed, as the {\it trans} conformer is less stable than
{\it cis} by about 2900~K, the thermal equilibrium ratio at 30~K is
about $ 10^{42}$! A discussion on the gas-phase chemical processes
that could lead to a large abundance of the {\it trans} conformer can
be found in \cite{neill12} and \cite{cole12}.

\section{Conclusion}

We have presented in this work the first set of collisional rate
coefficients for the rotational excitation of {\it cis}-methyl formate
by H$_2$ in the temperature range 5--30~K. The scattering calculations
were performed on the HCOOCH$_3$--He PES of
\cite{faure11}. Collisional propensity rules were shown to differ from
the radiative selection rules, with $\Delta J=2$ transitions being
favored. These collisional coefficients were employed in a non-LTE
radiative transfer treatment of methyl formate rotational lines and
the results were compared with the PRIMOS survey GBT observations
toward Sgr~B2(N). A total of 49 rotational transitions of methyl
formate, with upper levels below 25~K, were identified in the PRIMOS
data. Our best fit model suggests that these transitions sample a cold
($T_{\rm kin}\sim 30$~K), moderately dense ($n({\rm H_2})\sim
10^4$~cm$^{-3}$) and extended ($\sim$ arcminutes) region surrounding
Sgr~B2(N). In addition, our calculations indicate that all detected
emission lines with a frequency below 30~GHz are collisionally pumped
weak masers amplifying the background of Sgr~B2(N). This result
confirms the pioneering interpretation of \cite{brown75} based on the
1.61~GHz line and demonstrates the generality of the inversion
mechanism for the low-lying transitions of methyl formate. This
collisional process combined with a relatively strong and extended
continuum background could thus explain the emission spectra of many
complex organics. On the other hand, if the inversion cannot be
established (or is quenched) then absorption lines are expected. This
is indeed observed for the {\it cis}-methyl formate lines at 22.83~GHz
and for other species like e.g. {\it trans}-methyl formate for which
all observed features are in absorption \citep{neill12}. As a result,
emission spectra do not necessarily reflect higher kinetic
temperatures or densities but rather the possibility of population
inversions.

We have derived a column density for {\it cis}-methyl formate of $\sim
4\times 10^{14}$~cm$^{-2}$, in good agreement with previous estimates
based on millimetre-wave studies. This value is only a factor of
$\sim$10 larger than the column density deduced by \cite{neill12} for
the {\it trans} conformer in the same source. The origin of the
different conformers and isomers of methyl formate in low temperature
interstellar environments is still matter of debate \cite[see][and
  references therein]{cole12} but such a low {\it cis/trans} ratio
certainly provides a very strong constraint on chemical models
(provided that the two conformers are located in the same spatial
region). We note in this context the recent experimental evidence of
quantum tunneling in the gas-phase reaction between CH$_3$OH and OH
\citep{shannon13} suggesting that, in contrast to the usual
assumption, neutral-neutral reactions with an activation barrier could
play an important role in the cold post-evaporation gas-phase
chemistry.

It should be noted finally that collisional rate coefficients for
other heavy molecules are urgently needed in order to both accurately
determine the abundance and excitation of complex organics {\it via}
radiative transfer models and quantitatively investigate the
occurence, or absence, of maser action.

\section{Acknowledgements}
This work has been supported by the French INSU/CNRS Program
``Physique et Chimie du Milieu Interstellaire'' (PCMI) and by the
Agence Nationale de la Recherche (ANR-FORCOMS), contract
ANR-08-BLAN-022. The National Radio Astronomy Observatory is a
facility of the National Science Foundation operated under cooperative
agreement by Associated Universities, Inc. KS acknowledges support by
NSF grant CHE-1152899. Aurore Bacmann, Cecilia Ceccarelli, Pierre
Hily-Blant, S\'ebastien Maret and C\'eline Toubin are acknowledged for
useful discussions.

\bibliographystyle{apj}
\bibliography{faure0201}

\end{document}